\documentclass[aps,prl,twocolumn,superscriptaddress,longbibliography,nofootinbib]{revtex4-2}

\usepackage{lmodern} 
\usepackage[utf8]{inputenc} 
\usepackage[T1]{fontenc} 
\usepackage{microtype} 
\usepackage[english]{babel}
\usepackage[utf8]{inputenc}
\usepackage{color}
\usepackage{setspace}
\usepackage{graphicx}
\usepackage{amsmath}
\usepackage{amssymb}
\usepackage{slashed}
\usepackage[usenames,dvipsnames]{xcolor}
\usepackage[colorlinks=true, linkcolor=black, citecolor=ForestGreen]{hyperref}

\begin{document}

\title{Thermodynamic origin of the Landau instability of superfluids}
\author{Blaise Gout\'eraux}
\email{blaise.gouteraux@polytechnique.edu}
\affiliation{CPHT, CNRS, \'Ecole polytechnique, Institut Polytechnique de Paris, 91120 Palaiseau, France}
\author{Eric Mefford}
\email{ericmefford@uvic.ca}
\affiliation{Department of Physics and Astronomy, University of Victoria, Victoria, BC V8W 3P6, Canada}
\author{Filippo Sottovia}
\email{filippo.sottovia@polytechnique.edu}
\affiliation{CPHT, CNRS, \'Ecole polytechnique, Institut Polytechnique de Paris, 91120 Palaiseau, France}

\date{\today}

\preprint{CPHT-RR080.122022}

\begin{abstract}
In this work, we revisit the question of the linear stability of superfluid phases of matter. Famously, Landau predicted superfluid Helium would become unstable for large enough superfluid velocities. We demonstrate that this instability simply follows from a thermodynamic argument, by showing that its onset corresponds to a change of sign of one of the eigenvalues of the matrix of second derivatives of the free energy. Turning on dissipation and without any particular assumption on invariance under boosts, we show that a linear dynamical instability also develops, leading to exponential growth in time of perturbations around equilibrium. Specializing to Galilean superfluids and assuming the existence of quasiparticles, our criterion matches Landau's critical velocity. We also verify that it correctly reproduces the onset of the instability in relativistic superfluids constructed using gauge/gravity duality. Our work provides a simple, comprehensive and unified description of the Landau instability for superfluids independently of the microscopic details of the system.

\end{abstract}
\maketitle

\section{Introduction}

In its simplest incarnation, a superfluid phase of matter is formed when a global U(1) symmetry is spontaneously broken. Superfluids (and associated superconducting phases when the symmetry is local) are found across energy scales in many systems, such as Helium 4 and 3, \cite{landau1980course9}, quark matter, neutron stars \cite{Page:2006ud}, ultracold atomic gases \cite{pethick_smith_2008} as well as metals at low temperatures \cite{Leggett2006}.

Below the critical temperature, besides the conserved densities associated to symmetries of the system, the low energy effective theory includes an extra gapless mode $\varphi$, the Goldstone boson, \cite{chaikinlubensky1995}. Under U(1) transformations, the Goldstone transforms non-linearly $\varphi\mapsto\varphi+c$, and as a consequence only the superfluid velocity ${\bf v_s}\equiv\nabla\varphi$ is physical, not the phase itself. The superfluid velocity is topologically protected in the absence of free vortices (which are gapped at low temperatures), corresponding to an emergent higher-form conservation law for the winding of the superfluid phase $\int\star d\varphi$, \cite{Delacretaz:2019brr}. The existence of an emergent conservation law offers a new perspective on why the superfluid velocity also enters in the grand-canonical free energy, in addition to the temperature, the chemical potential and the normal fluid velocity, \cite{chaikinlubensky1995}. The overlap of this emergent conserved operator with the current leads to superfluidity.

Famously, Landau showed that superfluids are unstable for large enough values of the superflow, \cite{landau1980course9}. Upon using a Galilean transformation to boost the system to the superfluid rest frame, the energy of elementary quasiparticle excitations $\epsilon_q\mapsto\epsilon_q+{\bf q}\cdot {\bf v_s}$, with $q=|{\bf q}|$. For large enough ${\bf v_s}$ oriented anti-parallel to the wavevector ${\bf q}$, the quasiparticle energy in the superfluid rest frame becomes negative, leading to the creation of particles and so to loss of superfluidity. The critical velocity is found by solving the equation $\partial_q(\epsilon_q/q)=0$, given the quasiparticle energy $\epsilon_q$:
\begin{equation}
\label{LandauCrit}
v_L\equiv\textrm{min}_q\left(\epsilon_q/q \right).
\end{equation} 
This argument successfully predicts the critical velocity of Helium 4, where rotons become excited at large superfluid velocities and destroy superfluidity.\footnote{From the microscopic dispersion relation, rotons are seen to cause the Landau instability, but other critical velocity instabilities are seen experimentally and have been tied to vortex creation. For experiments designed explicitly to avoid such vortex creation, as in \cite{experiment1,experiment2}, the roton is indeed found to cause the Landau instability.}

While very intuitive, this argument does not easily generalize to systems without quasiparticles or without invariance under Galilean boosts. It also requires detailed knowledge of the microscopic excitations of the system. Relativistic superfluid phases \cite{ClarkPhD,putterman,Khalatnikov:106134,Bhattacharya:2011eea,Bhattacharya:2011tra,Schmitt:2014eka} are expected in the quark matter found in neutron stars or compact stars. They have been extensively investigated in the context of gauge/gravity duality applications to strongly-correlated condensed matter systems, with systems without long-lived quasiparticles such as high $T_c$ superconductors in mind, \cite{Gubser:2008px,Hartnoll:2008vx,Hartnoll:2008kx}. `Dirty' superfluids, where translations are explicitly broken, are also of interest. The main purpose of this work is to formulate a criterion for the Landau instability of superfluids which does not rely on the microscopic details of the system or invariance under boosts. 

Our main result is to show that the Landau instability is a local thermodynamic instability. Namely, we show that it coincides with the second derivative of the free energy with respect to the superfluid velocity becoming negative, $\partial^2 f/\partial v_s^2<0$. This thermodynamic instability is accompanied by a dynamical instability, signaled by the crossing of one of the gapless poles of the retarded Green's functions to the upper half complex frequency plane. We demonstrate this for a superfluid without any boost invariance and formally including dissipative gradient corrections. Next, we revisit the case of Galilean superfluids, showing that our criterion exactly matches Landau's original criterion. We show that this continues to hold for a relativistic superfluid, illustrating and explaining previous results obtained using gauge/gravity duality, \cite{Amado:2009ts,Lan:2020kwn}. Finally, we consider a `dirty' relativistic superfluid with a slowly-relaxing normal fluid velocity: there, the instability first appears in the thermal diffusion mode rather than in superfluid fourth sound. We give some technical details in a set of appendices, while the details of our holographic analyses will appear in a companion paper, \cite{Gouteraux:2023}. 

Some of these results appeared under different forms in previous literature. The connection between the Landau criterion and thermodynamic stability was pointed out for a Galilean superfluid at nonzero temperature in \cite{Andreev:2003,Andreev:2004}. The link between the Landau instability and a dynamical instability was discussed in \cite{Haber:2015exa,Andersson:2019ezz}, setting however the temperature to zero and only including a subset of dissipative terms. Here we substantially expand on these previous analyses and place them in a unified perspective.

Throughout this work, we adopt units where $\hbar=k_B=c=e=1$.

\section{Superfluid hydrodynamics}

The hydrodynamics of superfluids is well-known, \cite{landaubook,ClarkPhD,putterman,Khalatnikov:106134,chaikinlubensky1995,Bhattacharya:2011eea,Bhattacharya:2011tra,Schmitt:2014eka}. It is governed by the conservation equations following from invariance under time translations, space translations and U(1) global transformations:
\begin{equation}
\partial_t \epsilon +\partial_i j^i_\epsilon=0\,,\quad \partial_t g^i+\partial_j\tau^{ji}=0\,,\quad \partial_t n+\partial_i j^i=0\,,
\end{equation}
together with the Josephson relation $\partial_t\varphi+v_n^i\partial_i\varphi=-\mu$
which follows from gauge invariance. $\epsilon$, $n$, $g^i$ are the energy, charge and momentum densities, $j_\epsilon$ and $j$ the energy and charge currents, and $\tau$ is the spatial stress-tensor.

The thermodynamics of the system in the grand-canonical ensemble follows from the static partition function expressed as a local functional of the temperature $T$, the chemical potential $\mu$ and the norm of the superfluid velocity, $v_s=|{\bf v_s}|$. In order to facilitate the navigation between our different examples, we do not impose any particular boost symmetry at this point (see \cite{deBoer:2017ing,deBoer:2017abi,Novak:2019wqg,deBoer:2020xlc,Armas:2020mpr} for previous work on hydrodynamics of fluids without boosts) and work in the laboratory rest frame.
The first law of thermodynamics is
\begin{equation}
\label{1stlaw}
d\epsilon = T ds+\mu dn+{\bf v_{n}}\cdot d{\bf \pi}+{\bf h}\cdot d{\bf v_s}
\end{equation}
where $s$ is the entropy density, while ${\bf v_n}$ is the normal fluid velocity and ${\bf h}\equiv n_s ({\bf v_s}-{\bf v_n})$ is the conjugate quantity to the superfluid velocity. $n_s$ is the superfluid charge density, which quantifies the fraction of the density that participates in dissipationless superflow.

From \eqref{1stlaw}, \cite{chaikinlubensky1995}, we compute the divergence of the entropy current $T\partial_t s+T\partial_i(s v_n^i+\tilde j_q^i/T)\equiv\Delta$, where
\begin{equation}
\label{entropyproduction}
\Delta=-\tilde j_q^i\partial_i T/T-\tilde j^i\partial_i\mu-\tilde\tau^{ji}\partial_i v_{nj}-\tilde X\partial_i h^i
\end{equation}
 together with the constitutive relations for the currents
\begin{equation}
\label{constitutiverelations}
\begin{split}
&j^i=n v_n^i+h^i+\tilde j^i\,,\quad \tau^{ji}=p\delta^{ij}+v_n^i g^j+h^iv_s^j+\tilde\tau^{ji}\\
&j_\epsilon^i=(\epsilon+p)v_n^i-\partial_t\varphi h^i+\tilde j_q^i+\mu\tilde j^i+v_{nj}\tilde\tau^{ji}
\end{split}
\end{equation}
The pressure obeys the relation $p=-\epsilon+s T+n\mu+v_{n}^ig_i$, so that the first law can also be written $dp=sdT+n d\mu+g_idv_n^i-h_idv_s^i$. Symmetry under rotations implies that the stress tensor $\tau^{ij}$ is symmetric, and so $g^i=\rho v_n^i+h^i$, where $\rho$ is an undetermined function of all thermodynamic parameters.\footnote{Anticipating what follows, this can also be derived by noticing that $v_s^i=v_n^i+h^i/n_s$ and imposing that the matrix of static susceptibility is Onsager-symmetric.}
All tilded quantities are dissipative corrections to the ideal order constitutive relations. The Josephson relation is also corrected
\begin{equation}
\partial_t\varphi+v_n^i\cdot\partial_i\varphi=-\mu-\tilde X\,.
\end{equation}
At ideal order, $\Delta=0$, but in general, positivity of entropy production requires $\Delta\geq0$, which provides powerful constraints on the constitutive relations.

We now linearize the equations of motion around an equilibrium state characterized by background values of all thermodynamic quantities and associated sources, $(n,s,{\bf g},{\bf v_s})=(\bar n,\bar s,{\bf \bar g},{\bf\bar  v_s})+e^{-i\omega t+i{\bf q}\cdot x}(\delta n,\delta s,\delta{\bf g},\delta{\bf v_s})$ and $( \mu, T,{\bf  v_n}, {\bf  h})=(\bar \mu,\bar T,{\bf \bar v_n}, {\bf \bar h})+e^{-i\omega t+i{\bf q}\cdot x}(\delta \mu,\delta T,\delta{\bf  v_n}, \delta{\bf  h})$, which are related by a matrix of static susceptibilities, $\chi_{AB} = \delta O_A/\delta s_B=\delta^{2}W/\delta s_B\delta s_A$, defined as the variation of the vev $O_A$ of operator $A$ with respect to the source $s_B$ of operator $B$ holding other sources fixed, or equivalently the second variation of the static thermal free energy $W\equiv-T\log Z$ (where $Z$ is the static partition function). For the purposes of linearizing and solving the equations of motion, it is convenient to treat ${\bf v_s}$ as a vev and ${\bf h}$ as a source. However, from the perspective of the first law (and also for practical applications), it is more convenient to vary the superfluid velocity ${\bf v_s}$. We provide the correspondence between the thermodynamic derivatives in these two choices of ensemble in the appendix, denoting with a tilde static susceptibilities in the fixed ${\bf h}$ ensemble.

Upon linearizing and transforming to Fourier space, the equations of motion take the form 
\begin{equation}
\partial_t \delta O_A + \left(iq v_n\tilde\chi_{AB}+\tilde M_{AB}(q)\right)\delta s_B=0
\end{equation}
Due to the nonzero background normal velocity $v_n$, the fluctuations are dragged at a velocity $v_n$. Technically, this is an immediate consequence of the terms proportional to $v_n$ in the ideal constitutive relations. If the theory has boost invariance, we can boost to a frame where $v_n=0$. In the absence of boosts, we can simply consider changing to coordinates $x\to x-v_n t$ and redefining the frequency of perturbations to $\omega\to\tilde\omega=\omega-v_n q$. 
$\tilde M$ is a matrix which is expanded order by order in gradients $\tilde M(q)=iq \tilde M_1+q^2\tilde M_2+O(q^3)$, where ideal terms are contained in $\tilde M_1$ and dissipative terms at first order in gradients in $\tilde M_2$. The $\tilde M_i$ do not depend on $q$.\footnote{We always use the equations of motion at one order below to get rid of time derivatives in dissipative terms, which greatly simplifies writing expressions for the modes.} At ideal order, 
\begin{equation}
\label{Mideal}
\tilde M_1= \left(\begin{array}{cccc}
0&0&n&1\\
0&0&s&0\\
n&s&2(\rho-n_s)v_n+2 n_s v_s&v_s\\
1&0&v_s&0\end{array}\right)
\end{equation}
 which manifestly obeys Onsager relations (see Appendix).
 
The spectrum of collective modes is obtained by solving the equation $\det(-i\tilde\omega +\tilde M\cdot\tilde\chi^{-1})=0$. Upon increasing the superfluid velocity, an instability can only occur if one of the eigenvalues of the matrix $-i\tilde\omega +\tilde M\cdot\tilde\chi^{-1}$ becomes zero and then changes sign, i.e. $\det(M\cdot\tilde\chi^{-1})=\det \tilde M/\det\tilde\chi=0$. $\det \tilde M$ cannot vanish.  First, we observe that $\det \tilde M_1>0$. Then, we notice that $\tilde M_2$ is related to the quadratic form appearing in the divergence of the entropy current by $\Delta =q^2 \delta s_A\left(\tilde M_2\right)_{AB} \delta s_B$ (see Appendix). Imposing positivity of the quadratic form $\Delta\geq0$ then implies $\det \tilde M_2>0$. Thus, for an instability to occur, $\det\tilde \chi$ must diverge and change sign. Evaluating $\det\tilde\chi$, we see that this occurs when $\chi_{hh}\equiv1/\tilde\chi_{v_sv_s}=n_s-w \chi_{n_s h}=0$. In the Appendix, we show this in the limit when ${\bf v_n}$, ${\bf v_s}$ and the wavevector ${\bf q}$ are collinear, since the critical velocity is minimized in this limit.

We expect four collective modes (one for each independent fluctuation in the longitudinal sector) with a dispersion relation $\tilde\omega_i=v_i q -i \Gamma_i q^2+O(q^3)$. We can insert this expression in the determinant and solve it order by order in $q$. $\chi_{hh}=0$ implies that one of the modes has a vanishing velocity and attenuation, after which it crosses into the upper half plane. This leads to a perturbation growing exponentially with time and so to a dynamical instability.\footnote{ Here all velocities are generally nonzero, but it is not necessarily always the case. The argument goes through since the exponential growth in time is caused by the imaginary part changing sign.}

 The vanishing of $\chi_{hh}$ defines the critical superfluid velocity at which the instability develops:
\begin{equation}
\label{criticalvelocity}
\left.\partial_{v_s}\left((v_s-v_n)n_s\right)\right|_{v_s=v_s^c}=0
\end{equation}
This is our main result. We now proceed to demonstrate that it exactly matches the Landau criterion in Galilean superfluids, and correctly predicts the onset of the instability in holographic superfluids.

\section{Galilean superfluids and Landau's criterion}

To connect to the Landau criterion, we consider Galilean superfluids in the superfluid rest frame. In the Galilean limit, we impose that $g^i=j^i$ (setting the electron charge and the particle mass $e=m=1$). Going to the superfluid rest frame involves boosting from the lab frame to a frame moving with velocity $\bar{\bf v_s}$, parametrized by coordinates $t'=t$, ${\bf x'}={\bf x}-\bar{\bf v}_s t$ with $\partial_t' = \partial_t - \bar{v}_s^i\partial_i', \,\, \partial_i'=\partial_i$. From here on we drop the upper bar on ${\bf v_s}$.

The entropy-producing, dissipative corrections to the ideal order constitutive relations for Galilean superfluids were first written down in \cite{ClarkPhD, putterman} with an additional dissipative coefficient being identified in \cite{Bhattacharya:2011tra,Bhattacharya:2011eea,Banerjee:2016qxf}. The gradient corrections are invariant under Galilean boosts and so the conclusions drawn in the lab frame continue to hold in the rest frame up to a shift in the velocity, $\omega' = \tilde{\omega}-{v}_s q$. In particular, as we show in the Appendix, as $\chi_{hh}\to 0$ there is a mode in the rest frame with dispersion
\begin{align}
\omega_\star = (v_n - {v}_s)q + \chi_{hh}\left[\hat{v}q-i\hat{\Gamma}q^2+O(q^3)\right]
\end{align}
where $\hat{v}$ and $\hat{\Gamma}$ are non-zero constants. The linear dependence of the attenuation on $\chi_{hh}$ confirms that this mode becomes dynamically unstable as $\chi_{hh}$ changes sign. There is a subtlety in the superfluid rest frame, however. Using the rest frame identities $\mu_0 = \mu + {\bf v_s}\cdot{\bf v_n} - {\bf v_s}^2/2$, ${\bf g}_0 = n_n{\bf w}$, and ${\bf w} \equiv {\bf v_n} - {\bf v_s}$ the relative velocity, the thermodynamics is ignorant of $h$ and ${\bf v_s}$ \cite{landaubook}, e.g.
\begin{align}
dp = sdT + n d\mu_0 + {\bf g}_0\cdot d{\bf w}.
\end{align}
To reconcile this, we note that under a Galilean transformation $p\mapsto p$, but $\tilde{p}\mapsto \tilde{p} + {\bf h}\cdot{\bf v}_s = p$, or similarly, we identify $d\tilde{p} = dp$ for states with $\bar{{\bf v}}_s=0$. This identifies $\tilde{p}$ as the natural thermodynamic ensemble to compare states with the same $\bar{{\bf v}}_s$, as we should in the superfluid rest frame. The susceptibilities $\tilde{\chi}$ are constructed from variations of $\tilde{p}$ and in the Appendix, we show that every entry in $\tilde{\chi}$ diverges at the critical velocity. Hence, it suffices to consider a common microscopic condition for such a divergence.

In the superfluid rest frame, we write the pressure 
\begin{align} 
\tilde{p} = -T\int \frac{d^d q}{(2\pi)^d}\ln (1-e^{-(\epsilon_q-{\bf q}\cdot{\bf w})/T}).
\end{align}
$\epsilon_q(\mu_0, T)$ is the dispersion relation of the quasiparticle excitations, \cite{landau1980course9,Schmitt:2014eka}, which we expect to be a smooth function of its parameters, and we have boosted to a frame where the normal fluid composed of the thermal excitations moves relative to the superfluid at velocity ${\bf w}$. 

It is clear from the form of the pressure that as
\begin{align}
\label{galileanlandau}
{\bf w} \to {\bf w_c} = \text{min}_q \frac{\epsilon_q(\mu_0,T)}{q}
\end{align}
the susceptibilities will diverge, e.g.
\begin{align}
\lim_{w\to w_c}\tilde{\chi}_{ij} \propto \int \frac{d^d q}{(2\pi)^d}\frac{\mathcal{F}_{ij}(q,\mu_0,T)}{(\epsilon_q - {\bf q}\cdot{\bf w})^2}.
\end{align}
Here, $\mathcal{F}_{ij}$ depends on the particular susceptibility and smoothness of $\epsilon_q(T,\mu_0)$ implies that $\mathcal{F}$ is also smooth near ${\bf w}_c$. Hence, \eqref{galileanlandau} exactly reproduces the Landau criterion connecting instabilities of the macroscopic superfluid to microscopic excitations. In \cite{Andreev:2003,Andreev:2004}, the critical velocity for which $\chi_{hh}=0$ was related to the Landau critical velocity defined by \eqref{LandauCrit} in a different thermodynamic ensemble, but only found equality at zero temperature, assuming a phonon-roton quasiparticle dispersion relation. Here we see that we do not need to assume any specific dispersion relation.

\section{Relativistic superfluids}

In the relativistic limit, it is natural to express the conservation equations in covariant notation
\begin{equation} 
\nabla_\mu T^{\mu\nu}=0\,,\quad \nabla_\mu J^\mu=0\,.
\end{equation}
Invariance under Lorentz boosts imposes that the energy current is equal to the momentum $j_\epsilon=g$,  or equivalently that $T^{ti}=T^{it}$. It is convenient to work in the normal fluid rest frame, where at equilibrium the normal fluid velocity $u^{\mu}=(1,{\bf 0})$, and $u_\mu u^\mu=-1$. The constitutive relations, including first order derivative corrections, are well-known:
\begin{equation}
\label{relativisticconstitutive}
T^{\mu\nu}=\epsilon u^\mu u^\nu+p P^{\mu\nu}+2n_s\zeta^{(\mu}u^{\nu)}+\frac{n_s}{\mu}\zeta^\mu\zeta^\nu+\tilde T^{\mu\nu},
\end{equation}
\begin{equation}
J^\mu = n u^\mu + \frac{n_s}{\mu}\zeta^\mu + \tilde{J}^\mu\,,
\end{equation}
where we have defined the projector normal to the fluid velocity, $P^{\mu\nu}\equiv \eta^{\mu\nu}+u^\mu u^\nu$, restricting ourselves to a Minkowski background metric $\eta^{\mu\nu}$, and $\zeta^\mu\equiv P^{\mu\nu}\partial_\nu\varphi$ is the relative superfluid velocity up to a factor of $\mu$. Finally, the Josephon relation is
\begin{equation}
u^\mu\partial_\mu\varphi=-\mu-\tilde\mu\,.
\end{equation}
The first law of thermodynamics is $dp=sdT+nd\mu-n_s/(2\mu) d(\zeta^\mu\zeta_\mu)$, while the pressure and energy densities are related through $\epsilon+p=sT+n\mu$. 

We illustrate this using a gauge/gravity duality model of a superfluid \cite{Gubser:2008px,Hartnoll:2008vx,Hartnoll:2008kx} with finite background superflow \cite{Herzog:2010vz, Arean:2010wu,Bhattacharya:2011eea,Amado:2013aea}, which we outline in the Appendix and the details of which will appear in a companion paper \cite{Gouteraux:2023}. 

We construct the background black hole solution corresponding to a state with a non-trivial condensate and characterized by a nonzero temperature, chemical potential and superfluid velocity. Then we perturb around this state to obtain the spectrum of gapless, hydrodynamic modes, which correspond to the quasinormal modes of the black hole and which can be computed  using standard methods, \cite{Policastro:2002se,Son:2002sd,Policastro:2002tn,Herzog:2002fn}. These gapless modes were matched to the predictions of superfluid hydrodynamics at zero superfluid velocity in \cite{Amado:2009ts,Arean:2021tks}. At nonzero superflow, a complete classification of dissipative terms to first order in gradients appeared in \cite{Bhattacharya:2011eea,Bhattacharya:2011tra}. In a companion paper \cite{Gouteraux:2023}, we use a slightly different parametrization, equivalent to the `modified phase frame' of \cite{Bhattacharya:2011eea,Bhattacharya:2011tra}, which facilitates the comparison of the numerical dispersion relations to the hydrodynamic predictions.

\begin{figure}
\includegraphics[width=.4\textwidth]{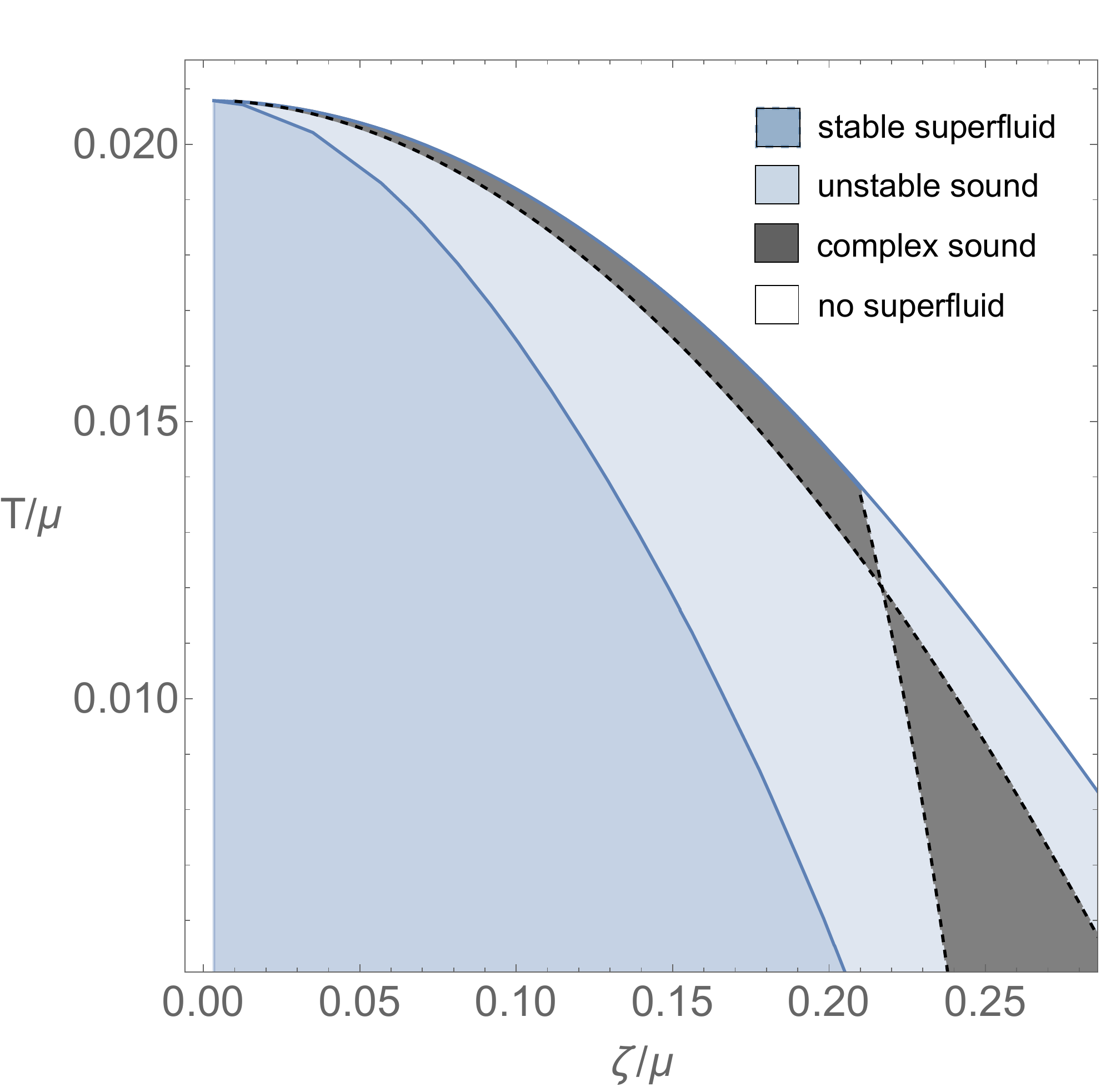}
\caption{\label{fig:relsup}Phase diagram for a holographic relativistic superfluid, with $\zeta=\sqrt{\zeta_\mu\zeta^\mu}$. Starting from the stable phase, the first instability appears as an unstable sound mode at the Landau critical velocity.}
\end{figure}

The phase diagram is depicted in figure \ref{fig:relsup}. We restrict ourselves to superfluid velocities anti-parallel to the wavevector, for which the instability is expected to arise for the lowest critical value of the superfluid velocity. At low temperatures and superfluid velocity, all hydrodynamic modes are stable. For fixed temperature, as the superfluid velocity increases, one of the sound modes crosses to the upper half plane, signaling the onset of an instability. The critical value for the superfluid velocity is precisely given by condition \eqref{criticalvelocity}.

There are also regions where two of the sound modes acquire complex velocities, which has been interpreted previously as a `two-stream' instability, \cite{Schmitt:2013nva,Haber:2015exa,Andersson:2019ezz}.

\section{Dirty superfluids}

Next, we study the instability when translations are explicitly broken, which is relevant for dirty superfluids. The momentum of the normal fluid relaxes at a rate $\Gamma_n\ll T$ which enters in the equations of motion as\footnote{A more careful analysis along the lines of \cite{Armas:2021vku} is warranted, but the extra dissipative transport coefficients there are subleading compared to the effect of $\Gamma_n$.}
\begin{equation}
\quad \partial_t g^i+\partial_j\tau^{ji}=-\Gamma_n v_n^i\,.
\end{equation}
For simplicity, we assume that the normal fluid has a vanishing background velocity, $\bar{\bf v_n}=0$. Then, the spectrum of collective excitations contains two sound modes (usually called fourth sound), one gapped mode $\omega=-i\Gamma_n+O(q)$ and a thermal diffusion mode
\begin{equation}
\omega=-i \frac{s^2\chi_{hh}}{(\chi_{sh}^2+\chi_{ss}\chi_{hh})\Gamma_n} q^2\,,
\end{equation}
which lies in the upper-half plane when $\chi_{hh}<0$. 

\begin{figure}
\includegraphics[width=.4\textwidth]{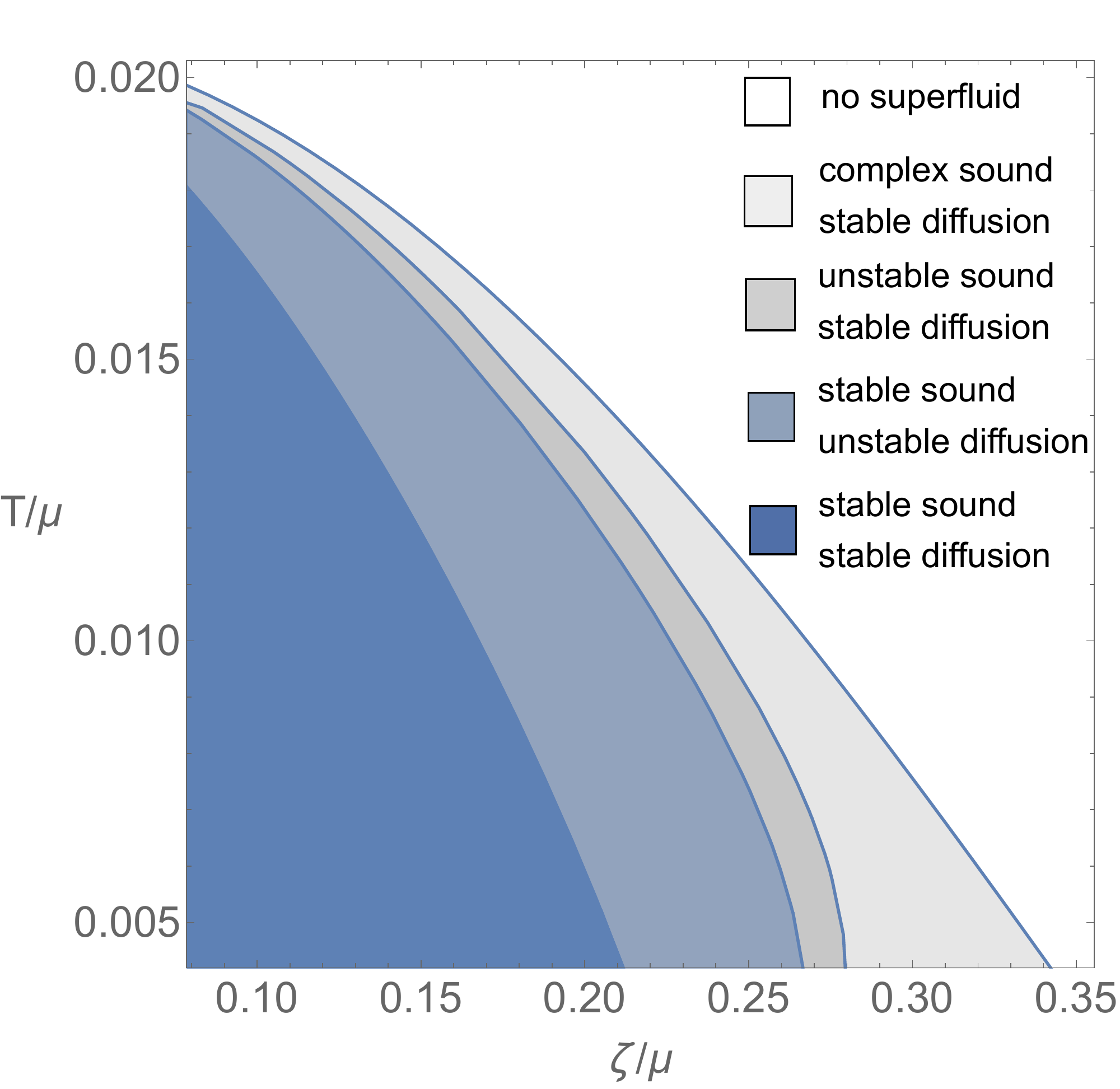}
\caption{\label{fig:brokentrans}Phase diagram for a holographic relativistic superfluid with weakly broken translations such that $\Gamma_n =\frac{1}{100}\frac{s}{4\pi(\mu n_n+sT)}\ll 1/T$, and $\zeta=\sqrt{\zeta_\mu\zeta^\mu}$. Starting from the stable phase, the first instability appears in the diffusion mode at the Landau critical velocity.}
\end{figure}

We illustrate this using a gauge/gravity duality model of superfluids with broken translations based on \cite{Andrade:2013gsa,Donos:2013eha} (see \cite{Andrade:2014xca,Ling:2014laa,Kim:2015dna, Gouteraux:2019kuy,Gouteraux:2020asq,Arean:2021tks} for previous investigations of such models).
Going through the same exercise as in the translation-invariant case, we produce the phase diagram in figure \ref{fig:brokentrans}. Our results confirm that the leading instability is given by the condition \eqref{criticalvelocity}, upon which the thermal diffusion mode crosses to the upper-half plane. As we further increase the superfluid velocity, this mode crosses back into the lower-half plane, and instead one of the sound modes becomes unstable. This happens when $\tilde\chi_{ss}=\chi_{ss}+\chi_{sh}^2/\chi_{hh}$ vanishes. This is allowed since in this region $\chi_{hh}<0$. Increasing further the superfluid velocity, both sound modes acquire complex velocities.

\section{Discussion and outlook}

In this work, we have demonstrated that superfluids are linearly dynamically unstable whenever the superfluid velocity becomes too large, and that this instability is of thermodynamic origin: one of the eigenvalues of the matrix of static susceptibility changes sign. Further, we have shown that this coincides with the Landau criterion for the critical velocity when we assume invariance under Galilean boosts, and the existence of quasiparticles. 

As advertised, our instability criterion does not rely on the microscopic details of the system, although of course these are implicitly contained in the specific dependence of the static partition function on thermodynamic parameters. It is equally valid whether the endpoint of the instability is the excitation of rotons or the nucleation of vortices.

A corollary of our analysis (under the simplifying assumption of collinearity) is that local thermodynamic stability together with positivity of entropy production is sufficient to guarantee linear dynamical stability: all hydrodynamic modes lie in the lower-half complex frequency plane. It would be interesting to prove this more generally.

\section{Acknowledgments}

\begin{acknowledgments}

We are grateful to Andreas Schmitt and Benjamin Withers for discussions. The work of B.~G.~ and F.~S.~ was supported by the European Research Council (ERC) under the European Union's Horizon 2020 research and innovation programme (grant agreement No758759). The work of E.~M.~was supported in part by NSERC and in part by the European Research Council (ERC) under the European Union's Horizon 2020 research and innovation programme (grant agreement No758759). We all gratefully acknowledge Nordita's hospitality during the Nordita program `Recent developments in strongly-correlated quantum matter' where part of this work was carried out.

\end{acknowledgments}

\bibliography{biblio}

 \newpage
 
\appendix

\section{\label{app} Appendix}

\section{Superfluid hydrodynamics in the laboratory frame}

In the lab frame, moving between ensembles with fixed ${\bf v_s}$ and fixed ${\bf h}$ is equivalent to the Legendre transformations
\begin{align} 
p \to \tilde{p} = p+{\bf h}\cdot{\bf v_s}, \quad \epsilon \to \tilde{\epsilon}-{\bf h}\cdot{\bf v_s}, \quad \epsilon+p \to \tilde{\epsilon}+\tilde{p}.
\end{align}

The matrix of static susceptibilities in the ensemble where ${\bf h}$ is held fixed in terms of static susceptibilities where ${\bf v_s}$ is held fixed is
\begin{widetext}
\begin{equation}
\tilde\chi = \left(\begin{array}{cccc}
\chi_{nn}+w\frac{\chi_{nh}\chi_{n_s n}}{\chi_{hh}} & \chi_{ns}+w\frac{\chi_{nh}\chi_{n_s s}}{\chi_{hh}}& \chi_{ng}+(n_s+w \chi_{n_s g})\frac{\chi_{nh}}{\chi_{hh}}&\frac{\chi_{nh}}{\chi_{hh}}\\
\chi_{sn}+w\chi_{n_s n}\frac{\chi_{sh}}{\chi_{hh}}&\chi_{ss}+w\frac{\chi_{n_s s}\chi_{sh}}{\chi_{hh}}& \chi_{sg}+(n_s+w \chi_{n_s g})\frac{\chi_{sh}}{\chi_{hh}}&\frac{\chi_{sh}}{\chi_{hh}}\\
v_n\left(\chi_{\rho n}+ w\frac{\chi_{n_s n}\chi_{\rho h}}{\chi_{hh}}\right)&v_n\left(\chi_{\rho s}+w\frac{\chi_{n_s s}\chi_{\rho h}}{\chi_{hh}}\right)&\rho +v_n\chi_{\rho g}+v_n(n_s+w\chi_{n_s g})\frac{\chi_{\rho h}}{\chi_{hh}}&1+v_n\frac{\chi_{\rho h}}{\chi_{hh}}\\
w\frac{\chi_{n_sn}}{\chi_{hh}}&w\frac{\chi_{n_ss}}{\chi_{hh}}&\frac{n_s+w\chi_{n_s g}}{\chi_{hh}}&\frac1{\chi_{hh}}
\end{array}
\right)
\end{equation}
\end{widetext}
To make our notation clear, $\chi_{nn}\equiv\left.\partial n/\partial \mu\right|_{T,v_n,v_s}$ while $\tilde\chi_{nn}\equiv\left.\partial n/\partial \mu\right|_{T,v_n,h}$. Similarly, $\chi_{nh}\equiv\left.\partial n/\partial v_s\right|_{\mu,T,v_n}$ or  $\chi_{n_sh}\equiv\left.\partial n_s/\partial v_s\right|_{\mu,T,v_n}$. We recall that $w=v_n-v_s$ is the relative velocity between the normal and superfluid components of the flow. We also note that
\begin{equation}
\chi_{hh}=\frac{\partial^2p}{\partial v_s^2} = \left(\frac{\partial^2\tilde{p}}{\partial h^2}\right)^{-1}=n_s-w\chi_{n_s h}
\end{equation}

The matrix $\tilde\chi$ should obey Onsager relations, which imply $\tilde\chi({\bf v_n},{\bf v_s},{\bf q})=\mathbb{S}\cdot\tilde\chi^{T}(-{\bf v_n},-{\bf v_s},-{\bf q})\cdot \mathbb{S}$ where $\mathbb{S}$ is a diagonal matrix with the eigenvalues of the vevs under time reversal, namely $(1,1,-1,-1)$. They give the following relations between static susceptibilities
\begin{widetext}
\begin{equation}
 \chi_{ns}=\chi_{sn}\,,\quad\chi_{ng}=v_n\chi_{\rho n}-\chi_{nh}\,,\quad\chi_{sg}=v_n\chi_{\rho s}-\chi_{sh}\,,\quad \chi_{n_sn }=\frac{\chi_{nh}}{w}\,,\quad \chi_{n_s s}=\frac{\chi_{sh}}{w}\,,\quad \chi_{n_s g}=\frac{v_n}w\chi_{\rho h}-\chi_{n_s h}\,.
\end{equation}
\end{widetext}
The matrix of retarded Green's functions is determined by
\begin{equation}
\label{defGR}
G^R=\tilde M\cdot\left(-i\omega \tilde\chi-\tilde M\right)^{-1}\cdot\tilde \chi\,.
\end{equation}
The Onsager relations imply
\begin{equation}
G^R({\bf v_n},{\bf v_s},{\bf q})=\mathbb{S}\cdot\left(G^R\right)^{T}(-{\bf v_n},-{\bf v_s},-{\bf q})\cdot \mathbb{S}\,.
\end{equation}
From the expression \eqref{defGR}, one can derive that the matrix $\tilde M$ also obeys Onsager relations:
\begin{equation}
\tilde M({\bf v_n},{\bf v_s},{\bf q})=\mathbb{S}\cdot \tilde M^{T}(-{\bf v_n},-{\bf v_s},-{\bf q})\cdot \mathbb{S}\,.
\end{equation}

\section{Dissipative superfluid hydrodynamics in the collinear limit}

In the main text, we work in the collinear limit with ${\bf v}_s||{\bf v}_n|| q$, since we expect the critical velocity to be minimized in this limit.\footnote{This assumption can be lifted. However, classifying dissipative corrections without assuming any boost invariance goes beyond the scope of this work. See \cite{Novak:2019wqg,Armas:2020mpr} for such an analysis in the case of fluids.} We start with the constitutive relations for the dissipative fluxes for boost agnostic superfluids. Defining $w = \sqrt{{\bf w}^2}$ and $\hat{w} = {\bf w}/w$, the constitutive relations can be expressed in terms of the scalars
\begin{align}
S_1 = \partial_i  h^i, \quad S_2 = \hat{w}^i \hat{w}^j \partial_i v_n^j, \quad S_3 =  \hat{w}^i \partial_i T, \quad S_4 = \hat{w}^i\partial_i \mu
\end{align}
In the notation of \eqref{constitutiverelations}
\begin{align}
\label{constitutiverelationsapp}
\tilde{j}& = -\hat{w}\left[\alpha_1 S_1 + \alpha_2S_2 + \alpha_3S_3+\sigma S_4\right]\,,\nonumber\\
\quad X &= -\left[\zeta_3S_1 + \zeta_1 S_2 + \zeta_2 S_3 + \alpha_1 S_4\right],\nonumber\\
 \tilde{j}^q &= -\hat{w}\left[\zeta_2 S_1+ \xi S_2 +\kappa S_3  + \alpha_3S_4\right], \nonumber\\
 \tilde{\tau}^{ij} &=-\hat{w}^i\hat{w}^j\left[ \zeta_1 S_1+\eta S_2+\xi S_3+\alpha_2S_4\right].
 \end{align}
 Here we have already used Onsager reciprocity to relate some transport coefficients in the constitutive relations.
 
The matrix $\tilde M$ can then be expanded $\tilde M(q) = iq\tilde M_1 + q^2\tilde M_2+O(q^3)$ with $\tilde M_1$ given by \eqref{Mideal} and $\tilde M_2$ by
\begin{align}
\tilde M_2 = \left(\begin{array}{cccc}\sigma&\alpha_3&\alpha_2&\alpha_1\\ \alpha_3&\kappa&\xi&\zeta_2\\ \alpha_2&\xi&\eta&\zeta_1\\\alpha_1&\zeta_2&\zeta_1&\zeta_3\end{array}\right), \quad \Delta = \sum_{i,j=1}^4 S_i \tilde M_2^{ij} S_j.
\end{align}
In the second expression, we have illustrated how the quadratic form appearing in entropy production \eqref{entropyproduction} is related to the $O(q^2)$ terms in the equations of motion. In particular, $\Delta\geq 0$ implies that $\tilde M_2$ is a positive semidefinite matrix. Since $\tilde M_2^T =\tilde  M_2$ thanks to Onsager relations, this is equivalent to the statement that $\det(\tilde M_2) \geq 0$. 

\section{Galilean limit}

Now, we illustrate the results in the Galilean invariant limit. Importantly, the fluxes in \eqref{constitutiverelations} are explicitly Galilean invariant. Through redefinitions of $\mu, T, {\bf v}_n$ at the first derivative order, we can fix that ${\bf g}$ receives no gradient corrections. In a Galilean invariant superfluid, boost invariance implies ${\bf j} = {\bf g}$ from which we conclude that in this frame $\alpha_1=\alpha_2=\alpha_3=\sigma=0$.  Solving for the hydrodynamic modes gives four sound modes with dispersion
\begin{align}
\omega_i(q) = (v_n+v_i) q- i\Gamma_i q^2+..., \quad i=1,2,3,\star.
\end{align}
where everything is collinear with $q$. Even in the collinear limit, the expressions for the velocities and attenuations are quite complicated but notably, as $\chi_{hh}\to 0$, the unstable sound mode, $\omega_\star$, simplifies. The velocity and attenuation are
\begin{widetext}
\begin{align}
v_\star &= \chi_{hh}\biggl(\frac{s^2/2}{n_nw^22\chi_{n_s s}^2+s\chi_{n_ss}(n-\chi_{n_s n}w^2)-\chi_{n_s n}s^2}\biggr), \nonumber\\
\Gamma_\star &= -\frac{\chi_{hh}}{4w}\biggl(s[n\chi_{n_s s}-s\chi_{n_sn}]+\chi_{n_ss}w^2[\chi_{n_ss}n_n-s\chi_{n_sn}]\biggr)^{-2}\gamma^A\tilde M_2^{AB}\gamma^B
\end{align}
\end{widetext}
where
\begin{align}
\gamma=\left(\begin{array}{c}
0\\
s[n_s-w^2\chi_{n_sn}]+n_n[s+2\chi_{n_ss}w^2]\\
-w s \chi_{n_ss}\\
sw(n\chi_{n_ss}-s\chi_{n_sn})
\end{array}\right)
\end{align}
Both $v_1$ and $\Gamma_1$ are Galilean invariant. From the form of $\tilde M_1$ and $\tilde M_2$, it is clear that if we boost to the rest frame of the superfluid, this mode has
\begin{align}
\omega_\star' = (v_n-{v}_s+v_\star)q - i\Gamma_\star q^2+...
\end{align}
with \emph{the same} $v_\star$ and $\Gamma_\star$. Finally, $\Gamma_\star$ is linearly dependent on $\chi_{hh}$ times a terms with \emph{definite sign} as follows from positivity of $\tilde M_2$. This demonstrates that as $\chi_{hh}$ passes through the origin such that $\chi_{hh}/w >0$, an instability appears. 

\section{Holographic model}

The holographic model that we consider is described on a manifold $(\Sigma, g)$ with boundary $(\partial \Sigma, \gamma)$ and described by the action,
\begin{widetext}
\begin{equation}
S = \int_\Sigma d^4x\sqrt{-g}\left[R+6-\frac{F^2}{4} - |D\Psi |^2 + 2|\Psi|^2 - \frac{1}{2}\sum_{i=1}^2(\partial \chi_i)^2 \right] + \int_{\partial\Sigma} d^3x\sqrt{-\gamma}\left(2\mathcal{K}-4 - |\Psi|^2 - R^{(\gamma)} + \frac{1}{2}\sum_{i=1}^2 (\partial\chi_{i})^2\right)
\end{equation}
\end{widetext}
where we work in units with $16\pi G_{4} = L_{AdS}=1$. The second term above is a boundary term that regulates ultraviolet (UV) divergences of the action and leads to a good variational principle \cite{Balasubramanian:1999re,Skenderis:2002wp}. Here, $\gamma_{\mu\nu}$ is the induced metric on $\partial\Sigma$ with associated Ricci scalar $R^{(\gamma)}$ and $\mathcal{K}$ the extrinsic curvature of the surface. $A_M dx^M$ is a $U(1)$ gauge field with associated field strength $F_{MN} = \partial_M A_N - \partial_N A_M$ and $\Psi$ is a complex scalar field charged under the $U(1)$ with covariant derivative $D_M\Psi = \partial_M \Psi - ieA_M\Psi$. The fields $\chi_i$ are massless scalars that break translational symmetry and lead to momentum relaxation \cite{Andrade:2013gsa,Andrade:2014xca}. The Ansatz $\chi_{i} = \alpha \delta_{ij}x^j$, where $j$ runs over the boundary spatial indices, retains the homogeneity of the background solution and leads to a momentum relaxation rate $\Gamma_n = \alpha^2 s/[4\pi(\mu n_n + sT)]$, \cite{Gouteraux:2019kuy,Gouteraux:2020asq}.  The equations of motion of this action are:
\begin{align}
0&=R_{MN}+\frac{1}{2}F_{MP}F^{P}_{\;\;N} - \frac{1}{2}\partial_M\chi_i \partial_N \chi_i - (D_M\Psi)(D_N\Psi)^*\nonumber\\
&\;\;\;\;+\frac{g_{MN}}{2}\left[|D\Psi|^2 -6 -2|\Psi|^2 + \frac{F^2}{4} + \frac{(\partial\chi_i)^2}{2}-R\right]\nonumber\\
0&=\frac{1}{\sqrt{-g}}\partial_M(\sqrt{-g}F^{MN}) - 2e^2|\Psi|^2 A^N\nonumber\\
0&=\frac{1}{\sqrt{-g}}\partial_M(\sqrt{-g}D^M\Psi) + ieA^MD_M\Psi +2\Psi\nonumber\\
0&=\frac{1}{\sqrt{-g}}\partial_M(\sqrt{-g}D^M\Psi)^* + ieA^MD_M\Psi^* +2\Psi^*\nonumber\\
0&=\frac{1}{\sqrt{-g}}\partial_M(\sqrt{-g}\partial^M\chi_i).
\end{align}
We choose an Ansatz for the metric and matter fields
\begin{align}
&ds^2 = -Ddt^2 + Bdr^2 + C_{xx}dx^2 + C_{xt}dt\,dx  + C_{yy}dy^2\nonumber\\
&A_Mdx^M = A_tdt + A_xdx, \quad \Psi = \Psi* = \psi
\end{align}
where all fields are a function of the radial coordinate only. In the UV $(r\to\infty)$,
\begin{align}
\label{UVconditions}
D(r) &= r^2 - \frac{\alpha^2}{4} - \frac{\epsilon}{3r}+O(r^{-2}),\nonumber\\
B(r) &=r^{-2},\nonumber\\
C_{xx}(r) &= r^2 + \frac{\alpha^2}{4}+ \frac{\epsilon-p}{3r}+O(r^{-2}),\nonumber\\
C_{xt}(r) &= \frac{\zeta n_s}{3r}+O(r^{-2}),\nonumber\\
C_{yy}(r)&= r^{2}+\frac{\alpha^2}{4}+\frac{p}{3r}+O(r^{-2}),\nonumber\\
A_t(r)&= \mu - \frac{n}{r}+O(r^{-2}),\nonumber\\
A_x(r) &= \zeta - \frac{\zeta n_s}{\mu}\frac{1}{r} + O(r^{-2}),\nonumber\\
\psi(r) &= \frac{\langle O^\psi\rangle}{2r^2} + O(r^{-3}).
\end{align}
Here, we have written the constants such that the conserved currents
\begin{align}
\langle T_{\mu\nu} \rangle = \frac{2}{\sqrt{-\gamma_{(b)}}}\frac{\delta S}{\delta \gamma_{(b)}^{\mu\nu}}, ,\quad \langle J^\mu \rangle = \frac{1}{\sqrt{-\gamma_{(b)}}}\frac{\delta S}{\delta A_{\mu,(b)}}, 
\end{align}
match \eqref{relativisticconstitutive} in equilibrium, as first shown in \cite{Sonner:2010yx}. Here the subscript $(b)$ means the diverging powers of $r$ are removed. We can use scale symmetries to fix the black hole horizon at $r=r_h$, where
\begin{align}
D &= B^{-1} = 4\pi T(r-r_h)+O(r-r_h)^2, \nonumber\\
 C_{xt} &= C_{xt}^h(r-r_h)+O(r-r_h)^2, \nonumber\\
 C_{xx} &= C_{xx}^h + O(r-r_h), \nonumber\\
 C_{yy} &= \frac{1}{C_{xx}^h}\left(\frac{s}{4\pi}\right)^2 + O(r-r_h), \nonumber\\
 A_t &= A_t^h(r-r_h) + O(r-r_h)^2,\nonumber\\
 A_x &= A_x^h + O(r-r_h),\nonumber\\
 \psi &= \psi_h + O(r-r_h).
\end{align}
The equations of motion lead to a conserved quantity when $\alpha = 0$, 
\begin{align}
\left[\sqrt{-g}\left(\frac{C_{yy}'}{C_{yy}D}+\frac{C_{yy}^2}{g}\left[\frac{\sqrt{-g}}{C_y^2D}\right]' - \frac{(A^2)'}{2D}\right)\right]'=\alpha^2\frac{\sqrt{-g}}{C_{yy}}.
\end{align}
Evaluating this in the UV and in the infrared for $\alpha = 0$ leads to 
\begin{align}
\epsilon+p = \mu n+sT,\quad 2p-\epsilon = -\frac{\zeta^2}{\mu}n_s.
\end{align}
We can then perturb the background solution with the following fields
\begin{align}
\delta g_{tt} &= Dh_{tt}(r) e^{-i(\omega t-qx)},\nonumber\\
 \delta g_{tx} &= \delta g_{xt} = C_{yy}h_{tx}(r) e^{-i(\omega t-qx)},\nonumber\\
 \delta g_{xx} &= C_{xx} h_{xx}(r) e^{-i(\omega t-qx)}, \nonumber\\
 \delta g_{yy} &= C_{yy} h_{yy}(r) e^{-i(\omega t-qx)}\nonumber\\
 \delta A_t &= a_t(r) e^{-i(\omega t-qx)},\nonumber\\
 \delta A_{x} &= a_x(r) e^{-i(\omega t-qx)},\nonumber\\
 \delta \Psi &= (\psi_r(r)+ i\psi_i(r))e^{-i(\omega t-qx)},\nonumber\\
 \delta \Psi* &= (\psi_r(r)- i\psi_i(r))e^{-i(\omega t-qx)},\nonumber\\
 \delta \chi_{x} &= \alpha \hat{\chi}(r) e^{-i(\omega t-qx)}
\end{align}
and all other fields vanishing. As explained in the main text, for simplicity, we only consider a wavevector that is collinear with the superfluid velocity. These fields are required to satisfy ingoing boundary conditions at the black hole horizon,
\begin{align}
h_{tt} &= (r-r_h)^{i\omega/4\pi T}\left[h_{tt}^h(r-r_h)+O(r-r_h)^2\right]\nonumber\\
h_{tx} &= (r-r_h)^{i\omega/4\pi T}\left[h_{tx}^h(r-r_h)+O(r-r_h)^2\right]\nonumber\\
h_{xx} &= (r-r_h)^{i\omega/4\pi T}\left[h_{xx}^h+O(r-r_h)\right]\nonumber\\
h_{yy} &= (r-r_h)^{i\omega/4\pi T}\left[h_{yy}^h+O(r-r_h)\right]\nonumber\\
a_{t} &= (r-r_h)^{i\omega/4\pi T}\left[a_t^h(r-r_h)+O(r-r_h)^2\right]\nonumber\\
a_{x} &= (r-r_h)^{i\omega/4\pi T}\left[a_{x}^h+O(r-r_h)\right]\nonumber\\
\psi_{r} &= (r-r_h)^{i\omega/4\pi T}\left[\psi_{r}^h+O(r-r_h)\right]\nonumber\\
\psi_{i} &= (r-r_h)^{i\omega/4\pi T}\left[\psi_{i}^h+O(r-r_h)\right]\nonumber\\
\hat{\chi} &=(r-r_h)^{i\omega/4\pi T}\left[\hat{\chi}^h+O(r-r_h)\right].
\end{align}
Solving the equations of motion for the fluctuations near the horizon, we find that four of the constants $\{h_{tt}^h, h_{tx}^h, h_{yy}^h, a_t^h\}$ are determined in terms of the others. In the UV, 
\begin{align}
h_{\mu\nu} &= h_{\mu\nu}^{(s)} + \frac{h_{\mu\nu}^{(2)}}{r^2} + \frac{h_{\mu\nu}^{(v)}}{r^3}+O(r^{-4}),\nonumber\\
a_\mu &= a_\mu^{(s)} + \frac{a_\mu^{(v)}}{r}+O(r^{-2}),\quad \psi_r = \frac{\psi_r^{(s)}}{r}+\frac{\psi_r^{(v)}}{r^2}+O(r^{-3})\nonumber\\
\psi_i &= \frac{\psi_i^{(s)}}{r}+\frac{\psi_i^{(v)}}{r^2}+O(r^{-3}),\quad \hat{\chi}_i = \hat{\chi}_i^{(s)} + \frac{\hat{\chi}_i^{(v)}}{r^2} + O(r^{-3}).
\end{align}
We interpret the constants $X^{(s)}$ above as fluctuations of the hydrodynamic sources and the constants $X^{(v)}$ as fluctuations of the hydrodynamic one point functions. Up to contact terms, the Green's function $G^{R}_{O^a O^b}(\omega,q) = -\delta\langle O^a\rangle /\delta s^b$. In the holographic prescription, poles of the Green's functions (which are not affected by contact terms) are given by solutions to the linearized equations of motion with $X^{(s)}=0$ \cite{Policastro:2002se,Policastro:2002tn,Kovtun:2005ev}--however, due to the horizon conditions, enforcing this boundary condition would overconstrain the problem. Instead, we fix four gauge invariant combinations to vanish corresponding to three constant diffeomorphisms $\beta_i = c_ie^{-i(\omega t-iqx)}$ and a $U(1)$ gauge transformation $\Lambda=\lambda e^{-(i\omega t-iqx)}$,
\begin{align}
\delta g_{\mu\nu} &= \nabla_\mu \beta_\nu + \nabla_\nu\beta_\mu, \quad \delta A_\mu = \pounds_\beta A_\mu + \partial_\mu \Lambda, \nonumber\\
\delta \chi_i &= \beta^\mu\partial_\mu \chi_i,\quad \delta \Psi = \beta^\mu\partial_\mu \psi + i\psi\Lambda,\nonumber\\
 \delta\Psi^* &= \beta^\mu\partial_\mu\psi - i\psi\Lambda.
\end{align}
For a given wavevector, $q$, the solutions to the linearized equations of motion, called quasinormal modes, exist only at certain frequencies $\omega(q)$. The frequencies with the smallest $|\omega(q)|$ will correspond to the dispersion relations of the hydrodynamic modes. In a companion paper, we confirm that these dispersion relations exactly match the dispersion relations of relativistic superfluid hydrodynamics (see \cite{Arean:2021tks} for a similar analysis without a background superfluid velocity).

The equations of motion for the background metric and matter fields as well as for the quasinormal modes must be solved numerically. Both are computed efficiently using pseudospectral methods over a Chebyshev grid \cite{Dias:2015nua}. For the background, we solve a boundary value problem using a Newton-Raphson relaxation algorithm with up to $N=400$ points. The boundary conditions we impose are the leading terms in the UV in \eqref{UVconditions} as well as regularity at the horizon. For the quasinormal modes, we solve a generalized eigenvalue problem over our numerical solutions using the boundary conditions previously discussed. Accurate results require higher than machine precision for the background which is efficiently implemented in Mathematica. The phase diagrams shown in figures~\ref{fig:relsup} and \ref{fig:brokentrans} were obtained both by calculating thermodynamic derivatives from the background solutions and via the quasinormal modes which are in agreement.

\end{document}